\documentclass[aps,prl,showpacs,floatfix,twocolumn,amsmath,amssymb,preprintnumbers]{revtex4-1}
\usepackage{mathrsfs}
\usepackage[figuresright]{rotating}
\usepackage{amsmath}
\usepackage{amssymb}
\usepackage{siunitx}
\usepackage{graphicx}
\usepackage{color}
\usepackage{dcolumn}
\usepackage{bm}
\usepackage[breaklinks=true,colorlinks=true,linkcolor=blue,urlcolor=blue,citecolor=blue]{hyperref}
\UseRawInputEncoding   

\usepackage{longtable}

\usepackage{multirow}
\usepackage{setspace}

\usepackage{soul}

\makeatletter

\newcommand{\Rmnum}[1]{\expandafter\@slowromancap\romannumeral #1@}
\makeatother

\begin{document}

\title{Ultrasonic investigation of the Kondo semimetal CeBi}

\author{Yupeng Pan$^{1}$}
\author{Xiaobo He$^{1}$}
\author{Shuo Zou$^{1}$}
\author{Hai Zeng$^{1}$}
\author{Yuqian Zhao$^{1}$}
\author{Ziyu Li$^{1}$}
\author{Yuesheng Li$^{1}$}
\author{Yongkang Luo$^{1}$}
\email[]{mpzslyk@gmail.com}
\address{$^1$Wuhan National High Magnetic Field Center and School of Physics, Huazhong University of Science and Technology, Wuhan 430074, China.}

\date{\today}

\begin{abstract}

We report the elastic properties of the Kondo semimetal CeBi by resonant ultrasound spectroscopy measurements at zero magnetic field. Clear elastic softening is found in bulk modulus $C_B$ below $\sim 60$ K. Such a softening in $C_B$, in addition to the anomalous temperature dependent Poisson's ratio, is hardly attributable to multipolar response for stable localized $4f$ orbital, but can be well described by a two-band model arising from the hybridization between conduction- and $4f$- electrons. These results probably are consequences of the valence fluctuations in this Kondo semimetal as originally suggested by a Fermi-surface expansion observed in a previous angle-resolved photoemission spectroscopy study [P. Li \textit{et al.}, Phys. Rev. B $\mathbf{100}$, 155110 (2019)].

\end{abstract}


\maketitle

\section{\Rmnum{1}. Introduction}

Materials encompassing valence instability have attracted a broad audience ever since decades ago \cite{Parks-ValenceInstab,Lawrence-ValFluc_RPP1981,Joyce-PuCoGa5XPS_PRL2003,Yuan-CeCu2Si2TwoSC,CaoG-SrVOFeAs,Zhai-Eu3Bi2S4F4}. It has been well known that many rare-earth (e.g., Ce, Eu, Yb) and actinide (e.g., U, Pu) compounds exhibit mixed-valence phenomena in the sense that the valences of \emph{quasilocalized} $f$ electrons are non-integers \cite{Misra-HFSystem}. The valence fluctuations in such systems are highly correlated and have bearing on many remarkable properties, like unconventional superconductivity\cite{Yuan-CeCu2Si2TwoSC,PuCoGa5_RUS_valencefluct_supercond}, (topological) Kondo insulators \cite{SmB6_1991,Kim-SmB6TKI_NM2014,Susaki-YbB12XPS_PRL1996}, etc. It is generally believed that Kondo, heavy-fermion, and mixed-valence behaviors correspond to different regimes in the so-called Kondo lattice system \cite{Misra-HFSystem}, depending on the strength of hybridization between conduction- and $f$-electrons, $J_{cf}$. Valence fluctuations are usually seen in the regime of strong $J_{cf}$ where the Kondo temperature $T_K$ is in the order of $10^2$ K (see e.g., SmB$_6$ \cite{SmB6_1991} and YbB$_{12}$ \cite{Susaki-YbB12XPS_PRL1996}); whereas in the Kondo regime (where $T_K$ is low), valence fluctuations behave as ``virtual" because each $f$ orbital is occupied by a single electron, \textit{viz}. the empty sites and doubly occupied sites essentially only act as virtual states \cite{Misra-HFSystem}.

Recently, CeBi was suggested as a rare example of valence fluctuations in the Kondo regime \cite{CeBi_ARPES_valence}. CeBi belongs to the NaCl-structured lanthanide monopnictides ($RePn$, $Re$ = rare-earth, $Pn$ = P, As, Sb, Bi). Kondo coupling is expected to be low in CeBi due to its semimetallic character with low carrier density ($\sim 0.03~e^{-}$/f.u. \cite{Suzuki-ReX}) \cite{Nozieres-EPJB1998,LuoY-CeNi2As2Pre}. Historically, CeBi has been extensively studied due to its rich magnetic phases and anomalous magnetotransport under field \cite{CeBi_magnetic_prorerty,Halg-CeBi_JMMM1982,PCCanfield2006crycgrow,Brinda2021crycgrow}. The recent discoveries of nontrivial topological electronic structures in $RePn$ \cite{Wu-LaBi_PRB2016,Nayak-LaBi_NC2017,He_PrBi_PRB2020,CeBi_ARPES_Nontrivaltopolog,CeBi_topologicaltransition_fermi,CeBi_Magnetotransport_topolog} re-sparked new interests in CeBi. Compared with other lanthanide monopnictides, one unique feature of CeBi is that a pronounced Fermi-surface expansion was recognized at low temperature by angle-resolved photoemission spectroscopy (ARPES) measurements, which was proposed as a signature of valence fluctuations of cerium \cite{CeBi_ARPES_valence}. However, owing to the small amount of this valence change ($\sim 0.01$), traditional transport or thermodynamic measurements hitherto have not been able to identify this feature.

In this paper, by employing the resonant ultrasound spectroscopy (RUS) technique that is both thermodynamic and symmetry-sensitive (cf Table \ref{Tab.1}), we successfully obtained the full set of elastic constants of CeBi. Most importantly, bulk modulus softening ($A_{1g}$ symmetry) and Poisson's ratio anomaly are observed below $\sim 60$ K, which probably are consequences of valence fluctuations in this Kondo semimetal. Other possible explanations based on structural or magnetic transition are also discussed.

\begin{table*}[!htbp]
\tabcolsep 0pt \caption{\label{Tab.1} Irreducible representation, strain, multipole, elastic constant, and illustration of strain in cubic system.}
\vspace*{0pt}
\renewcommand{\arraystretch}{1.8}
\def\temptablewidth{1.98\columnwidth}
{\rule{\temptablewidth}{0.5pt}}
\begin{tabular*}{\temptablewidth}{@{\extracolsep{\fill}}ccccc}
\hline
Symmetry   &  Strain $\varepsilon_{\Gamma}$ & Multipole $\hat{O}_{\Gamma}$  &    Elastic constant $C_{\Gamma}$  &   Deformation
\\ \hline
$\Gamma_{1}(A_{1g})$ & $\varepsilon_{B} =  \varepsilon_{xx} + \varepsilon_{yy} + \varepsilon_{zz}$   & $\hat{\mathbf{J}}^{2}$   &  $C_{B} = (C_{11} + 2C_{12})/3$  &  \multirow{2}{*}{\begin{minipage}{0.2\textwidth}\includegraphics[width=0.5\textwidth]{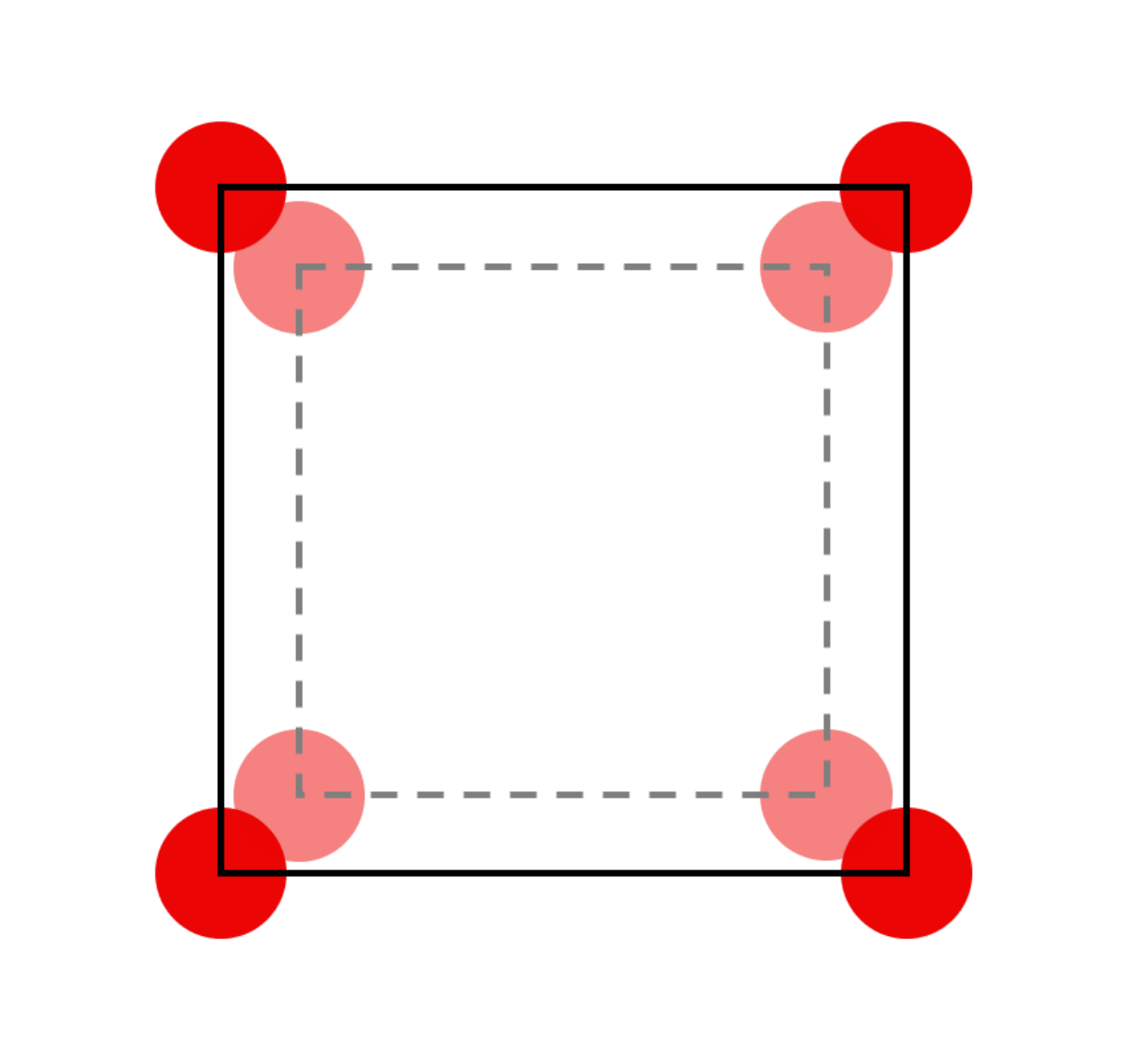}\end{minipage}}
\\
                     &             & $\hat{O}_4^0+5\hat{O}_4^4$
\\ \hline
$\Gamma_{3}(E_{g})$  & $\varepsilon_{u} =  (2\varepsilon_{zz} - \varepsilon_{xx} - \varepsilon_{yy})/\sqrt{6}$  & $\hat{O}_u\equiv3\hat{J}_{z}^2 - \hat{\mathbf{J}}^2$   &  $C_{T} = (C_{11} - C_{12})/2$  &  \multirow{2}{*}{\begin{minipage}{0.2\textwidth}\includegraphics[width=0.5\textwidth]{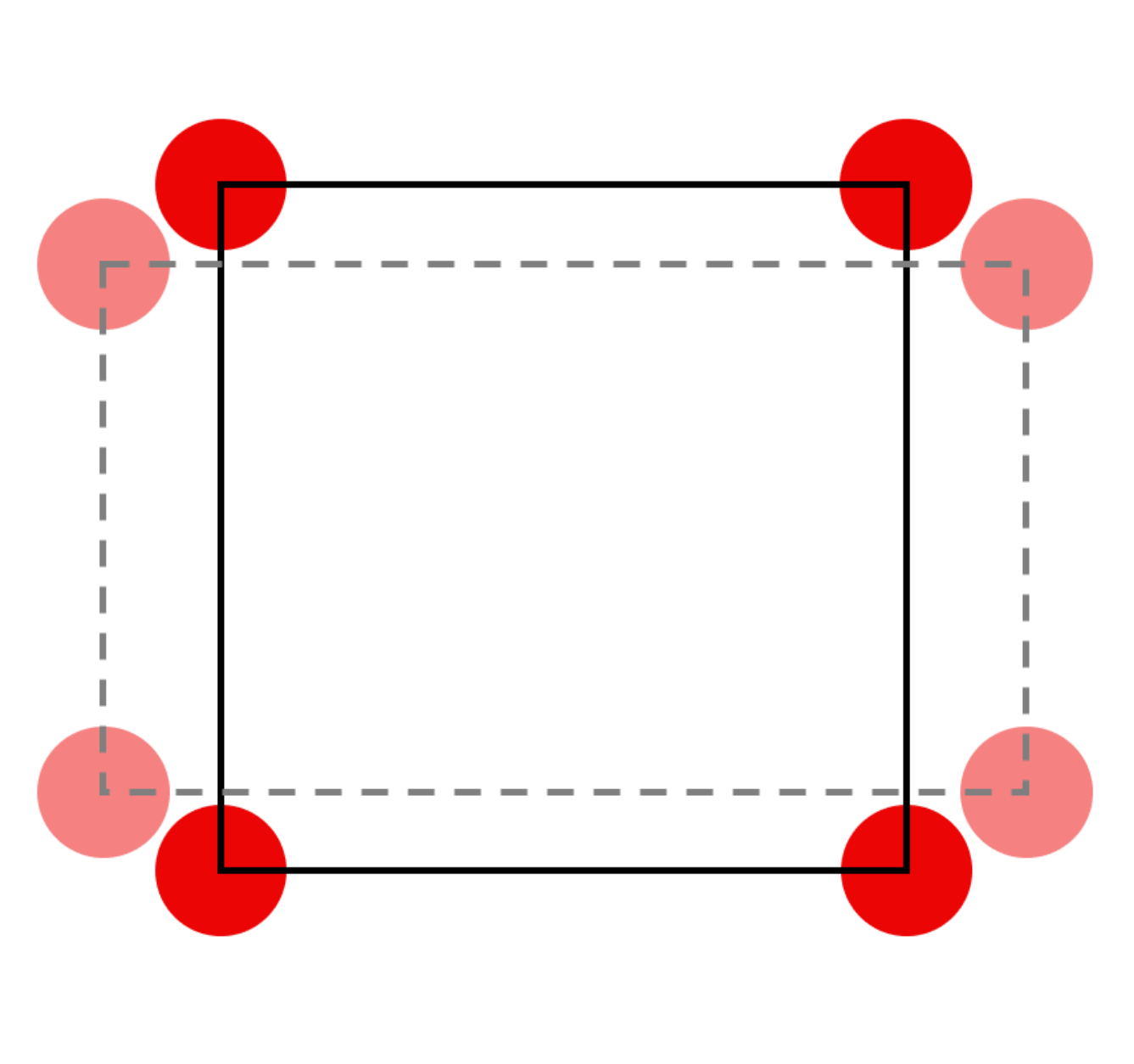}\end{minipage}}
\\
                 & $\varepsilon_{v} = (\varepsilon_{xx} - \varepsilon_{yy})/\sqrt{2}$  & $\hat{O}_v\equiv\hat{J}_{x}^2 - \hat{J}_{y}^2$   &  $C_{T} = (C_{11} - C_{12})/2$
\\ \hline
$\Gamma_{5}(T_{2g})$  & $\varepsilon_{yz}$  & $\hat{O}_{yz}$   &  $C_{44}$    & \multirow{3}{*}{\begin{minipage}{0.2\textwidth}\includegraphics[width=0.5\textwidth]{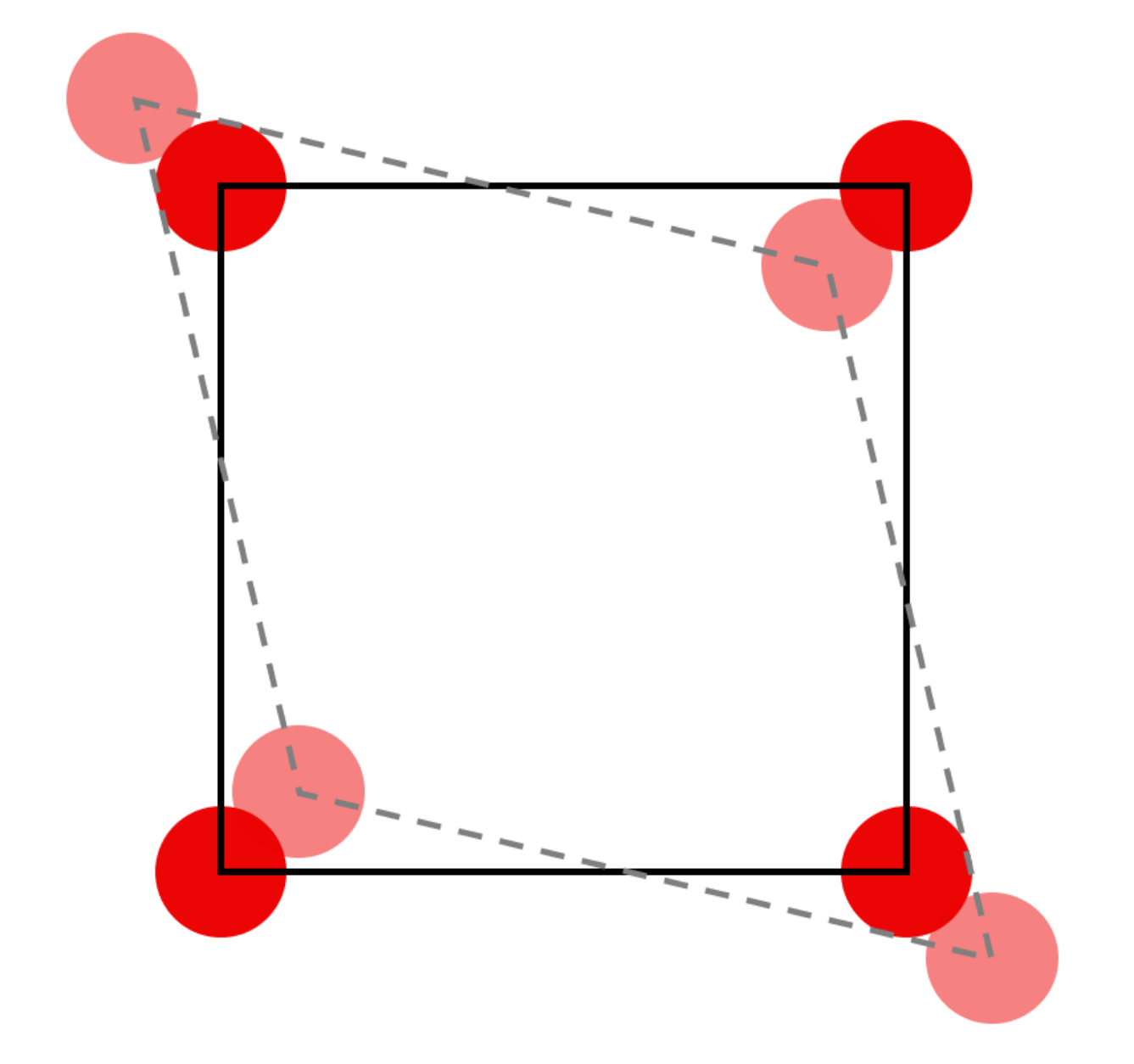}\end{minipage}}
\\
                      & $\varepsilon_{zx}$   & $\hat{O}_{zx}$   &  $C_{55}=C_{44}$
\\
                      & $\varepsilon_{xy}$  &  $\hat{O}_{xy}$  & $C_{66}=C_{44}$
\\
\hline
\end{tabular*}
{\rule{\temptablewidth}{0.5pt}}
\end{table*}

\begin{figure*}[!ht]
\vspace*{-10pt}
\hspace*{-10pt}
\includegraphics[width=18.0cm]{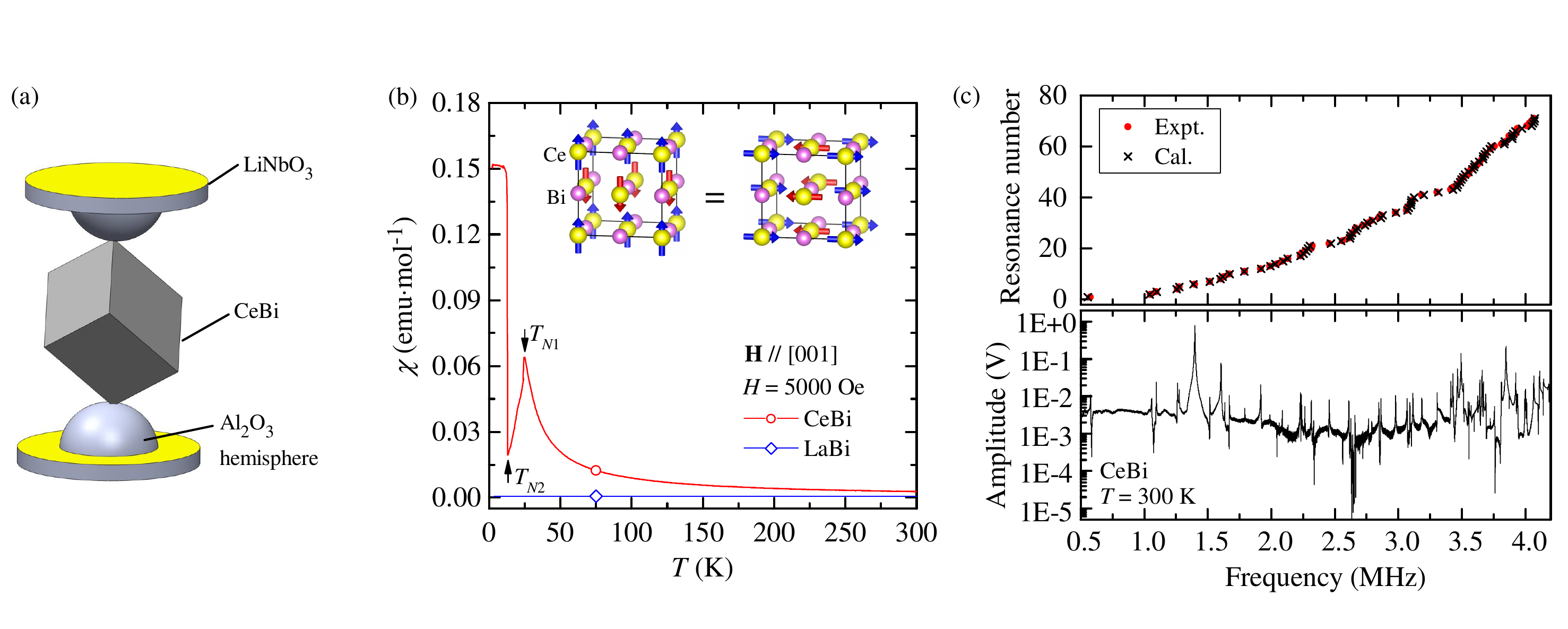}
\vspace*{-10pt}
\caption{(a) Scheme of RUS experimental set-up. (b) Magnetic susceptibility of CeBi and LaBi as a function of temperature ($\mathbf{H}\parallel[001]$). The data were taken in a field-cooling protocol. Two AFM transitions can be identified at $T_{N1}=25$ K and $T_{N2}=13.5$ K for CeBi. The magnetic susceptibility of LaBi is temperature-independently small, manifesting the absence of detectable ferromagnetic impurity. Inset, magnetic structure of CeBi for $T_{N2}<T<T_{N1}$ \cite{Halg-CeBi_JMMM1982}. (c) RUS spectrum of CeBi at 300 K. The upper panel shows the first 71 resonances, red dots - experimental data, and black crosses - calculated data.}
\label{Fig1}
\end{figure*}

\section{\Rmnum{2}. Experimental details}

High-quality single crystals of CeBi were grown by the Bi-self-flux method as mentioned in previous papers \cite{PCCanfield2006crycgrow,Brinda2021crycgrow}. Ce chunk (Alfa Aesar, 99.9$\%$) and Bi granule (Aladdin, 99.9999$\%$) were weighed in a molar ratio of 26:74. The mixture was transferred into an alumina crucible and sealed in an evacuated quartz tube. The quartz tube was heated up to 1200 $^{\circ}$C in 9 h, dwelt for 5 h, and slowly cooled down to 960 $^{\circ}$C at a rate of 2 $^{\circ}$C/h. At this temperature, the Bi flux was removed by centrifugation. The non-$4f$ reference compound LaBi was also synthesized by a similar method \cite{LaBi-crystalgrowm}. The chemical composition of the obtained compounds were confirmed by energy-dispersive x-ray spectroscopy (EDS). The average mole ratios are Ce : Bi = 51.1 : 48.9 in CeBi and La : Bi = 51.95 : 48.05 in LaBi, close to stoichiometry. The quality of crystallization and sample orientation were verified by Laue X-ray diffraction. More details can be found in \textbf{Supplemental Material (SM)} \cite{SM}. Magnetic susceptibility was measured by Magnetic Property Measurement System (MPMS-VSM, Quantum Design).

For RUS measurements, the CeBi sample was polished into a parallelepiped along the principle axes with dimensions 0.972$\times$0.790$\times$0.403 mm$^{3}$ and mass 2.54 mg. A similar parallelepiped was also prepared for LaBi, with dimensions 0.979$\times$0.856$\times$0.673 mm$^{3}$ and mass 4.45 mg. RUS measurements were carried out in a lock-in technique by sweeping frequency from 0.5 to 5 MHz at stabilized temperatures ranging from 5.4 to 300 K in a helium-flow cryostat (OptistatCF, Oxford). A cartoon illustration of the RUS setup is shown in Fig.~\ref{Fig1}(a). A pair of $Y$10-cut LiNbO$_3$-type transducers were used as ultrasound transmitter (bottom) and receiver (top), and Al$_{2}$O$_{3}$ hemispheres were employed for insulation and mechanical protection \cite{ultrasounddevice2020}. Additional details about RUS can be found in Ref.~\cite{Migliori-RUS,RA2019RUStoolbook,nature2014YBCO}.

\section{\Rmnum{3}. Results and Discussion}

\begin{figure*}[!ht]
\vspace{-0pt}
\hspace{-0pt}
\includegraphics[width=17.0cm]{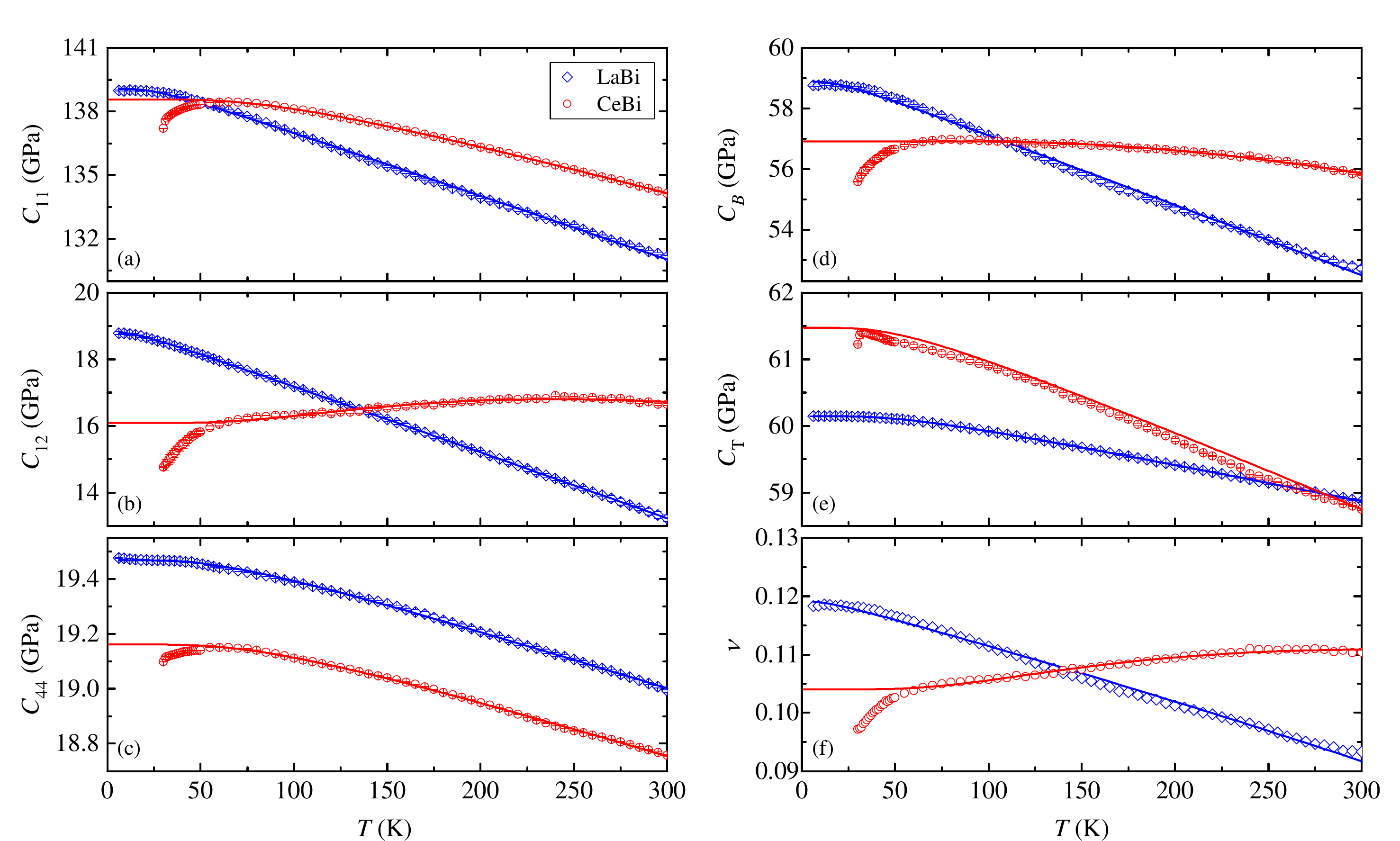}
\vspace{-0pt}
\caption{Temperature dependence of elastic constants of LaBi (blue diamonds) and CeBi (red circles). (a) $C_{11}$, (b) $C_{12}$, (c) $C_{44}$, (d) $C_{B} = (C_{11} + 2C_{12})/3$, and (e) $C_{T} = (C_{11} - C_{12})/2$. Panel (f) shows Poisson's ratio $\nu\equiv-S_{12}/S_{11}$. The solid lines are fits to Varshni's empirical description. }
\label{Fig2}
\end{figure*}

A representative vibrational spectrum of CeBi is shown in the bottom panel of Fig.~\ref{Fig1}(c). More spectra at other selected temperatures can be found in \textbf{SM} \cite{SM}. 71 resonant modes can be identified between 0.5 and 4 MHz, as shown in the top panel. Theoretically, by solving a three dimensional elastic wave function with given elastic constants, crystal geometry, and density, all the resonances can be uniquely deduced \cite{Migliori-RUS,Migliori-PhysicaB1993,Leisure-RUS}. Reversely, an abundant resonant frequencies can determine the elastic constants by a mathematical fitting process. This iteration continues until the root mean square (RMS) $R\equiv\sqrt{\{\sum_{n} [(f^n_{cal}-f_{exp}^n)/f^n_{exp}]^2\}/N}$ reaches minimum, where $N=71$ is the total number of identified peaks. In our analysis, RMS is less than 0.60\% for all the temperatures between 300 and 30 K. Due to the cubic symmetry of CeBi, the elastic modulus tensor ($\mathbf{C}_{6\times6}$) only has three independent elements, $C_{11}$, $C_{12}$ and $C_{44}$. The uncertainties of $C_{ij}$s were estimated by Levenberg-Marquardt Algorithm \cite{Migliori-RUS} as mentioned in \textbf{SM} \cite{SM}.

Before presenting the results of CeBi, we start from its non-$4f$ reference LaBi. The temperature dependence of $C_{11}$, $C_{12}$ and $C_{44}$ are displayed in Fig.~\ref{Fig2}(a-c). Further, bulk and shear moduli can be calculated by $C_{B} = (C_{11} + 2C_{12})/3$ and $C_{T} = (C_{11} - C_{12})/2$, respectively, and the results are shown in Fig.~\ref{Fig2}(d-e). All the elastic constants of LaBi increase monotonically upon cooling, and tend to level off at low temperature, as expected. No anomaly can be found in the full temperature window. The profiles of elastic constants can be well described by an empirical Varshni formula $C_{ij}^0(T)=a_{ij} - s/(e^{t/T}-1)$ \cite{varshni1970}, as signified by the blue solid lines in Fig.~\ref{Fig2}.

The elastic constants of CeBi are also shown in Fig.~\ref{Fig2} for comparison. It should be mentioned that for CeBi we only present the results for $T\geq29.8$ K, because we found all the resonant peaks become unresolvable when approaching $T_{N1}$ (cf Fig.~S2 in \textbf{SM} \cite{SM}). The reason for this is unclear. One possibility is that short-range antiferromagnetic (AFM) ordering forms near $T_{N1}$, and this scatters the propagating ultrasound wave randomly, therefore the resonant modes are severely dissipated. It is interesting to note that the resonant signal does not recover even for well below $T_{N1}$. Because the ordering is of A-type AFM structure in which ferromagnetic (001) planes with moments oriented perpendicularly to the plane are stacked antiferromagnetically in the sequence $+-+-$ \cite{Halg-CeBi_JMMM1982}, magnetic domains of different orientations but degenerate in energy are expected in this cubic crystalline structure [seeing inset to Fig.~\ref{Fig1}(b)]. We infer that the presence of domain walls will also attenuate ultrasound wave \cite{2022PRBSr2RuO4,UThBe1986domain}. In fact, such a multi-domain picture was also discussed in the ARPES work in Ref.~\cite{CeBi_topologicaltransition_fermi}. In addition, another possibility might be cubic-tetragonal structural distortion that accompanies the antiferromagnetic transition \cite{Hulliger-CeBi_struc}. This scenario was employed in a previous RUS work on MnV$_2$O$_4$ \cite{Luan_elasitictisis}.

For CeBi, all $C_{ij}$s stiffen initially upon cooling, and then soften at low temperature. After converting $C_{ij}$ into bulk and shear moduli, we find $C_T$ increases monotonically with decreasing temperature; whereas $C_B$ maximizes near 60 K, and then turns down rapidly [Fig.~\ref{Fig2}(d)]. Apparently, such a softening in $C_B$ should originate from the $4f$ character.

\begin{figure}[!ht]
\vspace{-0pt}
\hspace{0pt}
\includegraphics[width=8.5cm]{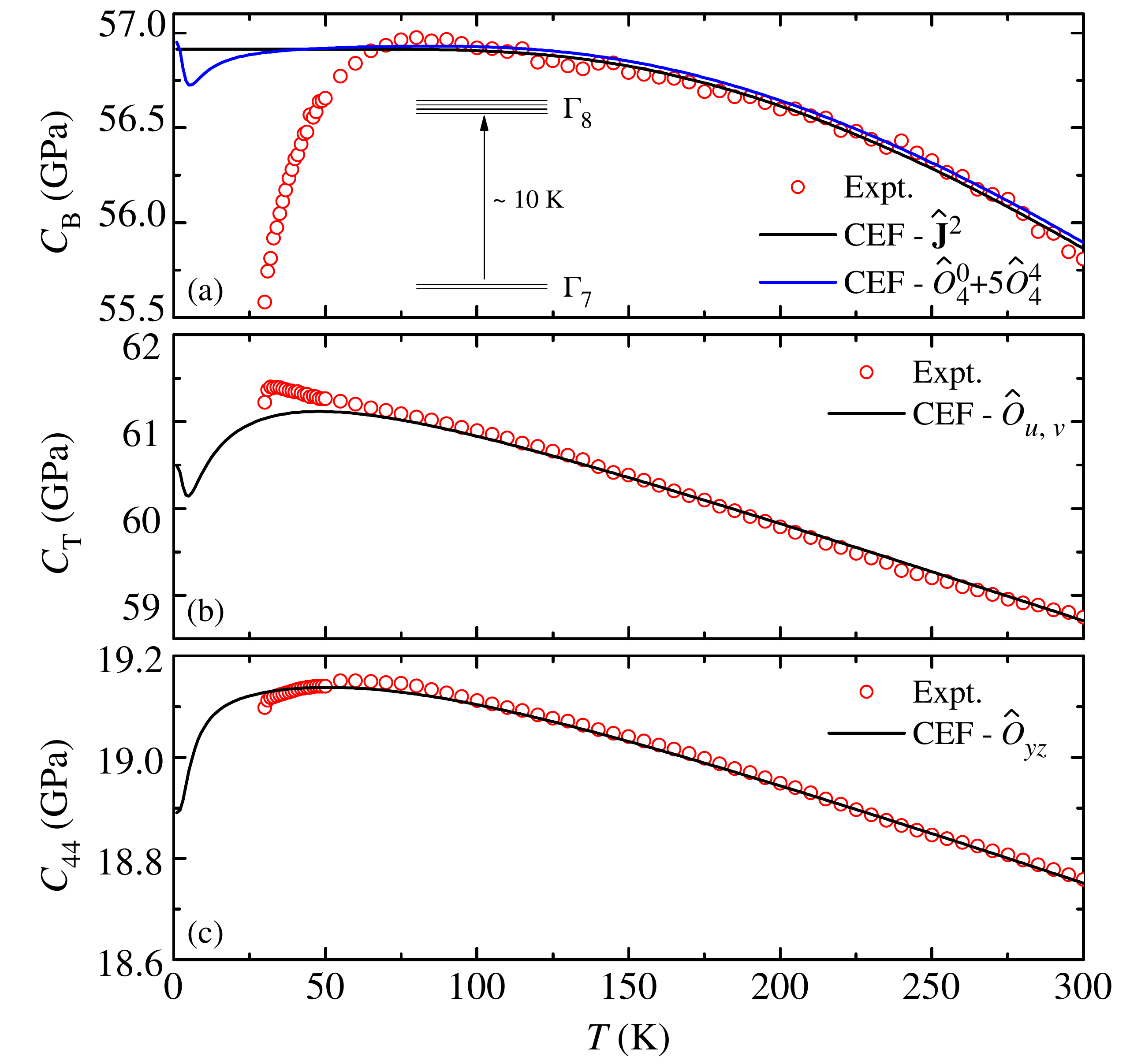}
\vspace{-10pt}
\caption{Temperature dependence of (a) $C_{B}$, (b) $C_{T}$ and (c) $C_{44}$ of CeBi. The solid lines represent the calculated elastic constants in the framework of multipole-strain coupling by assuming localized $4f$ orbital. The inset to panel (a) is a schematic diagram of CEF splitting of Ce$^{3+}$ $J=5/2$ multiplet.}
\label{Fig3}
\end{figure}

First, earlier elastic investigations on PrSb \cite{PrSbCEF1973,PrSbCEF1974} and other rare-earth contained compounds (e.g. Tb$_3$Ga$_5$O$_{12}$ \cite{Tb3Ga5O12CEF}) manifested that the  quadrupole-strain interaction can cause softening in certain elastic constants. To testify this idea, we performed these calculations based on the crystal electric field (CEF) effect with multipole of localized $4f$ electron. The total Hamiltonian is
\begin{equation}
\hat{H}=\hat{H}_{\text{CEF}}+\hat{H}_{\text{MS}}.
 \label{Eq1}
\end{equation}
The CEF Hamiltonian for Ce$^{3+}$ in $O_h$ point-group (cubic) symmetry is written as \cite{SSP1979CeBiCEF,PRB1988CeBiCEF}
\begin{equation}
\hat{H}_{\text{CEF}}=B_4^0 (\hat{O}_4^0 +5\hat{O}_4^4),
 \label{Eq2}
\end{equation}
where $\hat{O}_4^0$ and $\hat{O}_4^4$ are Stevens operators \cite{Stevens-Operators}, and $B_4^0$ is CEF parameter that can be determined experimentally. A previous neutron scattering experiment revealed that the $J=5/2$ multiplet of Ce$^{3+}$ ions splits into a ground doublet and a quartet that is $\sim10$ K above [cf inset to Fig.~\ref{Fig3}(a)] \cite{SSP1979CeBiCEF}. This yields $B_4^0\approx10/360$ K \cite{Stevens-Operators}. The second term of Eq.~(\ref{Eq1}) describes the multipole-strain coupling and takes the form of \cite{PrSbCEF1973,PrSbCEF1974}
\begin{equation}
\hat{H}_{\text{MS}} = -g_{\Gamma_{\lambda}}\hat{O}_{\Gamma_{\lambda}}\varepsilon_{\Gamma_{\lambda}}.
 \label{Eq3}
\end{equation}
Here $\Gamma_{\lambda}$ ($\lambda = 1,3,5$) are irreducible representations, and $g_{\Gamma_\lambda}$ are the coupling constants. The irreducible representation and the associated strain and elastic constants are summarized in Table~\ref{Tab.1}. Our subsequent analysis will be focused on the symmetrized elastic constants $C_{B}$, $C_{T}$, and $C_{44}$. The calculated results are shown as the solid lines in Fig.~\ref{Fig3}. Although this simulation can reproduce $C_{44}(T)$ satisfactorily, it fails in $C_B$ and $C_T$. In particular, the calculated strain susceptibility of $\Gamma_{1}(A_{1g})$ is almost zero for all temperatures, so the quadrupole-strain interaction of $\Gamma_{1}(A_{1g})$ has no contribution to $C_B$. The situation does not get too much improved even if we upgrade the quadrupole ($\hat{\mathbf{J}}^2$) to hexadecapole ($\hat{O}_4^0+5\hat{O}_4^0$) that also couples to $\varepsilon_B$ \cite{Tb3Ga5O12CEF,YbB12_2021}, seeing the blue solid line in Fig.~\ref{Fig3}(a). In short, these discrepancies suggest that the approach by only taking into account the multipolar response for \emph{stable} $4f$ orbital hardly copes with the elastic softening in CeBi below $\sim 60$ K.

Second, previous ultrasound-velocity measurements on SrTiO$_3$ revealed that bulk and shear moduli soften sharply when approaching the structural transition at 112 K, while the onset of this change occurs $\sim 20$ K prior to the structural distortion \cite{Bell-SrTiO3}. Furthermore, the presence of structural domains near the transition also causes the failure of ultrasound measurements, similar to the situation we confront with in CeBi. According to the low-temperature XRD measurements by Hulliger, CeBi undergoes a cubic-tetragonal distortion right at $T_{N1}$ \cite{Hulliger-CeBi_struc}. It, therefore, is reminiscent of structural-transition-induced elastic softening in CeBi. Besides structural transitions, magnetic transitions can also cause softening in certain symmetry channels. A representative example is CoF$_2$ that becomes an antiferromagnet below 39 K, while remarkable softening (mainly in shear modulus) appears as high as $\sim$90 K \cite{Co2FJPCM2014}. For both structural and magnetic transitions, in order to cause such elastic softening well above the transition temperature, it is necessary that dynamical local ordering couples to the acoustic modes. To further demonstrate or exclude these possibilities, additional microscopic and dynamic measurements are in demand.

\begin{figure}[!ht]
\vspace{-0pt}
\hspace{-20pt}
\includegraphics[width=9.5cm]{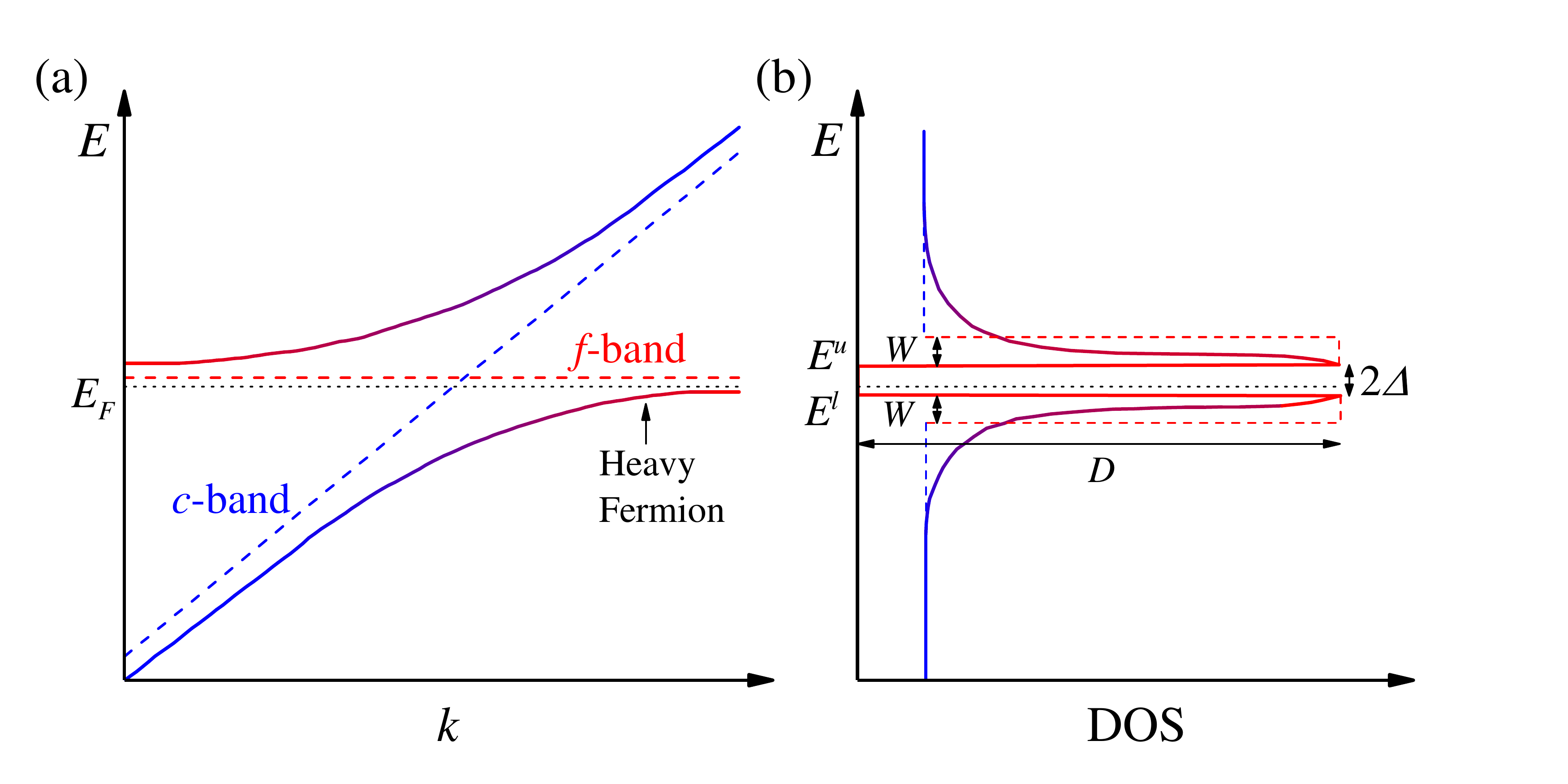}
\vspace{-10pt}
\caption{(a) Sketch of the $c-f$ hybridization. (b) A gap ($2\Delta$) opens near $E_F$ that leads to the two bands. For simplicity, the energy dispersion of the renormalized bands is ignored (see the rectangular-shaped DOS). Noteworthy that in reality, CeBi is a semimetal with a small but finite DOS at $E_F$, which might be contributed by other bands. }
\label{Fig4}
\end{figure}

An alternative possible explanation for the elastic anomaly in CeBi might be valence fluctuations. This is usually understood in terms of a two-band model as depicted in Fig.~\ref{Fig4}. Due to the $c-f$ hybridization, the quasiparticle bands renormalize to form the upper band ($E^u$) and lower band ($E^l$) that are separated by the hybridization gap ($2\Delta$). The interband hopping is realized by introducing $\hat{c}^\dag_{\mathbf{k},u(l)}$ and $\hat{c}_{\mathbf{k},u(l)}$, the creation and annihilation operators of an electron in the band u(l) with wave vector $\mathbf{k}$. In this model, the multipole-strain interaction is replaced by\cite{Kurihara-Pnictide,YbB12_2021}
\begin{equation}
\hat{H}_{\text{MS}}=-g_{\Gamma_\lambda}\sum_{\mathbf{k}}\left(
  \begin{aligned}
    \begin{array}{c}
       \hat{c}^\dag_{\mathbf{k},u} \\
       \hat{c}^\dag_{\mathbf{k},l} \\
    \end{array}
  \end{aligned}
\right)^T \left(
  \begin{aligned}
    \begin{array}{cc}
       d^{uu}_{\mathbf{k},\Gamma_{\lambda}} & d^{ul}_{\mathbf{k},\Gamma_{\lambda}} \\
       d^{lu}_{\mathbf{k},\Gamma_{\lambda}} & d^{ll}_{\mathbf{k},\Gamma_{\lambda}} \\
    \end{array}
  \end{aligned}
\right)\left(
  \begin{aligned}
    \begin{array}{c}
       \hat{c}_{\mathbf{k},u} \\
       \hat{c}_{\mathbf{k},l} \\
    \end{array}
  \end{aligned}
\right)\varepsilon_{\Gamma_{\lambda}},
\label{Eq4}
\end{equation}
where $d^{ij}_{\mathbf{k},\Gamma_{\lambda}}=\langle i|\hat{O}_{\Gamma_{\lambda}}|j\rangle$ ($i,j=u,l$). As the changes in cerium valence (Ce$^{3+}$ $\leftrightarrow$ Ce$^{4+}$) is also accompanied with an expansion / shrinkage in ionic radius, it is natural that $C_B$ which couples to an isotropic strain with $\Gamma_1$ ($A_{1g}$) symmetry will response. This scenario has been successfully applied for the valence fluctuations in many Kondo lattice compounds like SmB$_6$ \cite{SmB6_1991}, CeNiSn \cite{CePdSn_CeNiSn_1991AFM}, YbB$_{12}$ \cite{YbB12_2021} etc. To be more specific, quantitative analysis is made as following.

\begin{table*}[!ht]
\tabcolsep 0pt \caption{\label{Tab.2} Fitting parameters determined by the two-band model [Eq.~(\ref{Eq7})] calculation of elastic constants $C_{B}$, $C_{T}$ and $C_{44}$ of CeBi.}
\vspace*{0pt}
\renewcommand{\arraystretch}{1.8}
\def\temptablewidth{2\columnwidth}
{\rule{\temptablewidth}{0.5pt}}
\begin{tabular*}{\temptablewidth}{@{\extracolsep{\fill}}cccccccc}
\hline
 $C_{\Gamma_\lambda}$  &  $a_{\Gamma_\lambda}$ (GPa)   & $s$ (GPa)   &  $t$ (K)  & $2\Delta$ (K)   & $W$ (K)  & $Dg^2_{\Gamma_\lambda}(d_{\Gamma_\lambda}^{uu}-d_{\Gamma_\lambda}^{ll})^2$ (10$^9$J/m$^3$)  
\\ \hline
$C_{B}$ & 57.1(2)    & 10.1(1)   & 710.9(1)  & 2.6(4)  &  8.1(1)  &   39.7(4)  
\\
$C_{T}$  & 61.5(2)  & 1.7(1)   &  151.0(1)  & 2.6(4)  &  8.1(1)  & 13.2(5)    
\\
$C_{44}$   & 19.1(1)  & 0.5(3) &  248.3(3)   & 2.6(4)  &  8.1(1)  & 2.0(2)    
\\
$C_B$ (SmB$_6$ \cite{SmB6_1991})  &  &  &    &  160  &  150  & 3.3  
\\
$C_B$ (YbB$_{12}$ \cite{YbB12_2021}) &  &  &     &  140 &  55 &  25.6  
\\
\hline
\end{tabular*}
{\rule{\temptablewidth}{0.5pt}}
\end{table*}

\begin{figure}[!ht]
\vspace{-5pt}
\hspace{0pt}
\includegraphics[width=8.5cm]{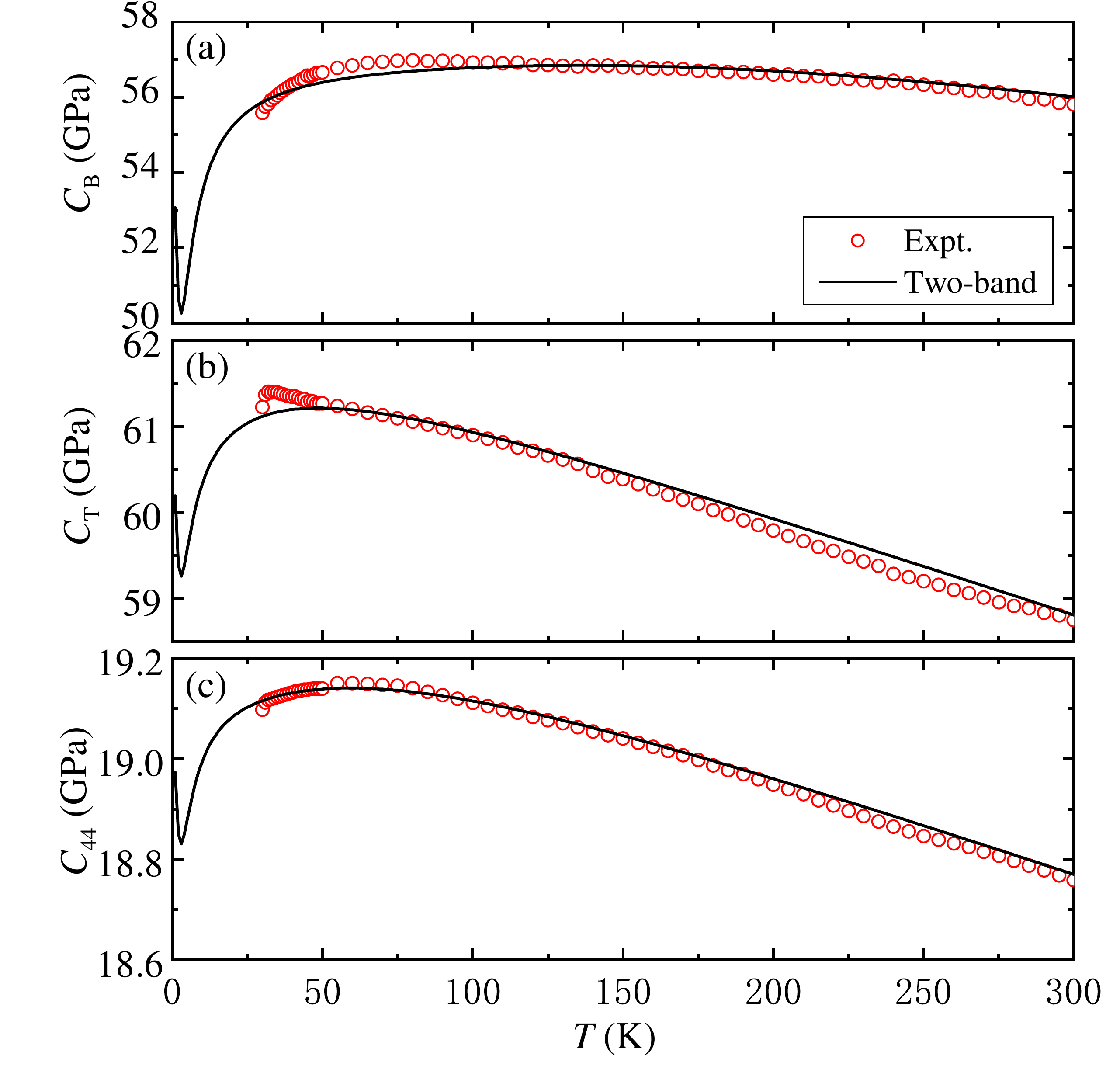}
\vspace{-15pt}
\caption{Temperature dependence of (a) $C_{B}$, (b) $C_{T}$ and (c) $C_{44}$. The solid lines indicate the fits to the two-band model.}
\label{Fig5}
\end{figure}

The free energy of this deformed system reads \cite{YbB12_2021,SmB6_1991}:
\begin{equation}
\begin{split}
F = & \frac{1}{2}C_{\Gamma_\lambda}^0\varepsilon_{\Gamma_\lambda}^2 + nE_F\left(\varepsilon_{\Gamma_\lambda}\right) \\
&-k_BT\sum_{i(=u,l),\mathbf{k}}\ln\left\{1+\exp\left[-\frac{E_{\mathbf{k}}^i\left(\varepsilon_{\Gamma_\lambda}\right)-E_{F}\left(\varepsilon_{\Gamma_\lambda}\right)}{k_BT}\right]\right\},
\end{split}
 \label{Eq5}
\end{equation}
in which $n$ is the total number of conduction electrons, $E^{u(l)}_\mathbf{k}(\varepsilon_{\Gamma_\lambda})$ is the upper- (lower-) band energy in the presence of strain\cite{YbB12_2021,SmB6_1991},
\begin{subequations}
\begin{align}
E^u_{\mathbf{k}}(\varepsilon_{\Gamma_{\lambda}})=E^u_{0,\mathbf{k}}-g_{\Gamma_\lambda}d^{uu}_{\mathbf{k},\Gamma_{\lambda}}\varepsilon_{\Gamma_\lambda}+
\frac{g_{\Gamma_\lambda}^2|d^{ul}_{\mathbf{k},\Gamma_{\lambda}}|^2}{2\Delta}\varepsilon^2_{\Gamma_\lambda},\label{Eq.6a}\\
E^l_{\mathbf{k}}(\varepsilon_{\Gamma_{\lambda}})=E^l_{0,\mathbf{k}}-g_{\Gamma_\lambda}d^{ll}_{\mathbf{k},\Gamma_{\lambda}}\varepsilon_{\Gamma_\lambda}-\frac{g_{\Gamma_\lambda}^2|d^{ul}_{\mathbf{k},\Gamma_{\lambda}}|^2}{2\Delta}\varepsilon^2_{\Gamma_\lambda}.\label{Eq.6b}
\end{align}
\end{subequations}
After some tedious mathematics, the elastic constants can be deduced\cite{SmB6_1991,YbB12_2021},
\begin{equation}
\begin{split}
C_{\Gamma_\lambda}(T) \equiv &\frac{\partial^2F}{\partial\varepsilon^2_{\Gamma_\lambda}} = C_{\Gamma_\lambda}^0(T)-\frac{1}{4}Dg^2_{\Gamma_\lambda}\left(d_{\Gamma_\lambda}^{uu}-d_{\Gamma_\lambda}^{ll}\right)^2\\
&\times\left[\tanh\left(\frac{\Delta+W}{2k_BT}\right)-\tanh\left(\frac{\Delta}{2k_BT}\right)\right]\\
&+Dg^2_{\Gamma_\lambda}\left|d^{ul}_{\Gamma_\lambda}\right|^2\frac{2k_BT}{\Delta}\ln\left(\frac{\cosh\left(\frac{\Delta}{2k_BT}\right)}{\cosh\left(\frac{\Delta+W}{2k_BT}\right
)}\right).
\end{split}
\label{Eq7}
\end{equation}
Here we have ignored the $\mathbf{k}$ dependence in the calculations, and the energy dispersion of density of states (DOS) of the renormalized bands is also simplified by treating them as constant ($D$), seeing Fig.~\ref{Fig4}. In this process, $C^0_{\Gamma_\lambda}(T)$ is assumed to obey Varshni's form; $\Delta$, $W$, $g^2_{\Gamma_\lambda}D(d^{uu}_{\Gamma_\lambda}-d^{ll}_{\Gamma_\lambda})^2$, and $g^2_{\Gamma_\lambda}D|d^{ul}_{\Gamma_\lambda}|^2$ are the fitting parameters. In reality, the third term of Eq.~(\ref{Eq7}) turns out to be non-crucial, while the other fitting parameters are summarized in Table~\ref{Tab.2}, and the derived temperature dependent $C_B$, $C_T$ and $C_{44}$ are displayed in Fig.~\ref{Fig5}. The elastic softening below $\sim 60$ K can be well reproduced by this model, and this strongly suggests the presence of valence fluctuations in this Kondo semimetal. The fitting yields the hybridization gap $2\Delta=2.6$ K and renormalized band width $W=8.1$ K. The small $2\Delta$ is consistent with the weak Kondo coupling as expected in this low-carrier-density Kondo semimetal \cite{Suzuki-ReX}. Furthermore, we notice that the fitting for $C_T$ is not as good as for $C_B$ and $C_{44}$, probably because of the simplified two-band model as well as some other factors e.g. AFM transition that have not been taken into account in this fitting.

Finally, it should be noted that the valence fluctuations in CeBi can be further evidenced by Poisson's ratio ($\nu$), seeing Fig.~\ref{Fig2}(f). Poisson's ratio is a measure of the expansion of a material in directions perpendicular to the specific direction of compressing. In this context, $\nu$ is defined as $-S_{12}/S_{11}$, where $S_{11}$ and $S_{12}$ are the elements of the elastic compliance tensor $\mathbf{S}_{6\times6}\equiv\mathbf{C}^{-1}$. For most conventional metals and ionic crystals, $0.25\lesssim\nu<0.35$ \cite{Greaves-Poisson_NM2011}, while for most covalent systems $\nu\sim0.1$ \cite{Shein-FeAs_elastic,PanY-ScZrNbTaRhPd}, as are the cases for both LaBi and CeBi. Most importantly, $\nu$ of CeBi gradually decreases below $\sim 250$ K, and the slope accelerates below $\sim 60$ K, mimicking the behavior of $C_{12}(T)$. Such an anomalous $T$ dependence of $\nu$ is also indicative of valence fluctuations. In a mixed-valence system, uniaxial compression can dictate the nearly-degenerate valence orbitals to adopt the configuration with smaller ionic radius (e.g., Ce$^{4+}$ in this context) by promoting the $c-f$ hybridization, and this likely reduces the extent of expansion at the perpendicular direction. Correspondingly, Poisson's ratio will be suppressed (even to negative values in some cases) \cite{PuCoGa5_RUS_valencefluct_supercond}. In CeBi, $\nu$ decreases by only $\sim$0.008 from 60 to 30 K, which probably suggests a small valence change of Ce. Although we do not have the data for below 30 K, considering the small hybridization gap we have obtained, this conclusion potentially remains valid even for low temperature. Qualitatively, this is in agreement with the small amount of valence change $\sim 0.01$ estimated by ARPES measurements \cite{CeBi_ARPES_valence}.

\section{\Rmnum{4}. Conclusions}

In conclusion, the full set of symmetry-resolved elastic constants of the Kondo semimetal CeBi are investigated by resonant ultrasound spectroscopy measurements at zero magnetic field. Below $\sim$ 60 K, clear softening behavior is observed in elastic constants especially in bulk modulus $C_B$. Poisson's ratio decreases abnormally below $\sim$ 60 K, too. Such anomalies are absent in the non-$4f$ reference LaBi. These peculiar features can not be understood by a simple multipole-strain coupling with stable localized $4f$ orbital, but can be well reproduced by a two-band model arising from the $c-f$ hybridization. The possible explanations based on valence fluctuations, structural or magnetic transitions are discussed for these elastic anomalies. Additional microscopic and dynamic measurements should be invited to further clarify this issue.

\section{Acknowledgments}

The authors thank Yifeng Yang for helpful discussions. This work is supported by National Key R\&D Program of China (2022YFA1602602), the open research fund of Songshan Lake Materials Laboratory (2022SLABFN27), Guangdong Basic and Applied Basic Research Foundation (2022B1515120020), and National Natural Science Foundation of China (12274153).


%

\newpage

\renewcommand{\thefigure}{S\arabic{figure}}
\renewcommand{\thetable}{S\arabic{table}}
\renewcommand{\theequation}{S\arabic{equation}}
\onecolumngrid

\newpage

\begin{center}
{\bf \large
{\it Supplemental Material:}\\
Ultrasonic investigation of the Kondo semimetal CeBi
}
\end{center}

\setcounter{table}{0}
\setcounter{figure}{0}
\setcounter{equation}{0}
\setcounter{page}{1}

\small
\begin{center}
Yupeng Pan$^{1}$, Xiaobo He$^{1}$, Shuo Zou$^{1}$, Hai Zeng$^{1}$, Yuqian Zhao$^{1}$, Ziyu Li$^{1}$, Yuesheng Li$^{1}$, and Yongkang Luo$^{1*}$\email{mpzslyk@gmail.com}\\
$^1${\it Wuhan National High Magnetic Field Center and School of Physics, Huazhong University of Science and Technology, Wuhan 430074, China.}\\
\date{\today}
\end{center}
\normalsize
\vspace*{15pt}

\section{SM I. S\lowercase{ample-quality characterization}}
The chemical compositions of CeBi and LaBi were determined by energy-dispersive x-ray spectroscopy (EDS) measurements. In CeBi, the average mole ratio Ce : Bi = 51.1 : 48.9, while in LaBi, the ratio La : Bi = 51.95 : 48.05; both are close to the stoichiometric ratio.

\begin{figure*}[!ht]
\vspace*{-60pt}
\includegraphics[width=17cm]{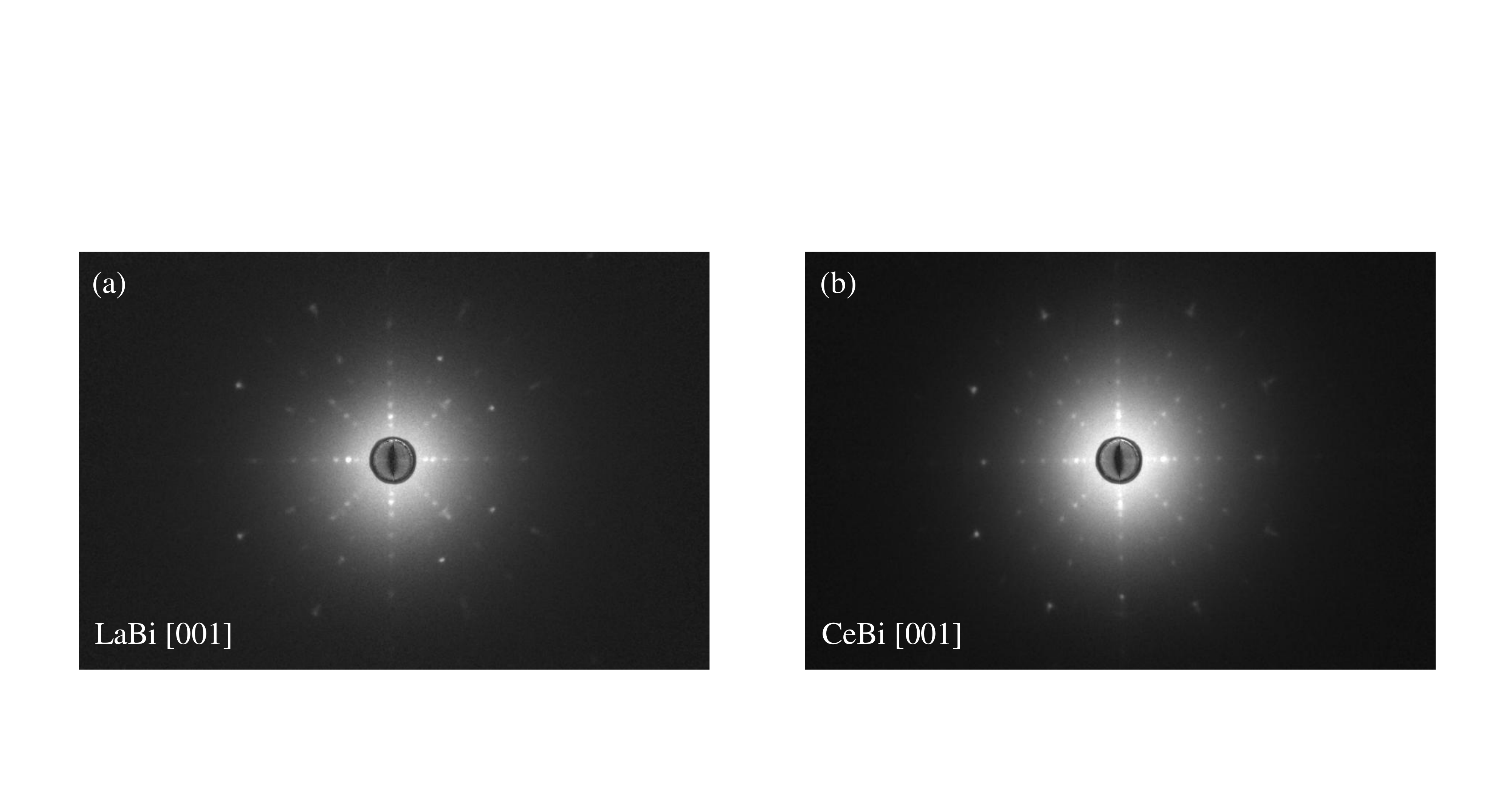}
\vspace*{-30pt}
\caption{\label{FigS1} Laue XRD pattern of LaBi (a) and CeBi (b) for polished samples on [001] surfaces.}
\end{figure*}

The sample quality of the polished single crystals was also characterized by Laue X-ray diffraction (XRD). The nice patterns on [001] surface as shown in Fig.~\ref{FigS1} demonstrate the high-quality of crystallization. No  signature of multi-grain or impurity can be seen. The measurements also confirm the nearly-perfect orientation of the polished samples, the extent of misalignment less than $\sim 1^{\circ}$.

\section{SM II. E\lowercase{lastic constants in cubic symmetry}}
In an elastic three-dimensional material, we can write its Hooke's law:
\begin{equation}
    \bm{\sigma} = \mathbf{C} \cdot \bm{\varepsilon}.
    \label{EqS1}
\end{equation}

In detail, we can also write this equation in three dimensions. However, this description is cumbersome. Usually the so-called Voigt notation is employed to simplify the subscripts:
\begin{equation}
  xx \leftrightarrow 1;~~ yy \leftrightarrow 2;~~ zz \leftrightarrow 3;~~ yz, zy \leftrightarrow 4;~~ xz, zx \leftrightarrow 5;~~ xy, yx \leftrightarrow 6.
  \label{EqS2}
\end{equation}

In this representation, the Hooke's law is transformed to a 6th-order tensor:

\begin{equation}
\left(
  \begin{aligned}
    \begin{array}{c}
       \sigma_{1}\\
       \sigma_{2}\\
       \sigma_{3}\\
       \sigma_{4}\\
       \sigma_{5}\\
       \sigma_{6}\\
    \end{array}
  \end{aligned}
\right)
=
\left(
\begin{aligned}
    \begin{array}{cccccc}
       C_{11} & C_{12} & C_{13} & C_{14} & C_{15} & C_{16}\\
       C_{21} & C_{22} & C_{23} & C_{24} & C_{25} & C_{26}\\
       C_{31} & C_{32} & C_{33} & C_{34} & C_{35} & C_{36}\\
       C_{41} & C_{42} & C_{43} & C_{44} & C_{45} & C_{46}\\
       C_{51} & C_{52} & C_{53} & C_{54} & C_{55} & C_{56}\\
       C_{61} & C_{62} & C_{63} & C_{64} & C_{65} & C_{66}\\
    \end{array}
  \end{aligned}
\right)
\left(
  \begin{aligned}
    \begin{array}{c}
       \varepsilon_{1}\\
       \varepsilon_{2}\\
       \varepsilon_{3}\\
       \varepsilon_{4}\\
       \varepsilon_{5}\\
       \varepsilon_{6}\\

    \end{array}
  \end{aligned}
\right).
    \label{EqS3}
\end{equation}

The elastic constants obey the primary symmetry $C_{ij} = C_{ji}$, so there are at most 21 different elastic constants to describe a system. This amount can be further reduced by symmetry \cite{Migliori-RUS}. In a cubic symmetrical system, Eq. (\ref{EqS3}) can be written as:

\begin{equation}
\left(
  \begin{aligned}
    \begin{array}{c}
       \sigma_{1}\\
       \sigma_{2}\\
       \sigma_{3}\\
       \sigma_{4}\\
       \sigma_{5}\\
       \sigma_{6}\\
    \end{array}
  \end{aligned}
\right)
=
\left(
\begin{aligned}
    \begin{array}{cccccc}
       C_{11} & C_{12} & C_{12} & 0 & 0 & 0\\
       C_{12} & C_{11} & C_{12} & 0 & 0 & 0\\
       C_{12} & C_{12} & C_{11} & 0 & 0 & 0\\
       0 & 0 & 0 & C_{44} & 0 & 0\\
       0 & 0 & 0 & 0 & C_{44} & 0\\
       0 & 0 & 0 & 0 & 0 & C_{44}\\
    \end{array}
  \end{aligned}
\right)
\left(
  \begin{aligned}
    \begin{array}{c}
       \varepsilon_{1}\\
       \varepsilon_{2}\\
       \varepsilon_{3}\\
       \varepsilon_{4}\\
       \varepsilon_{5}\\
       \varepsilon_{6}\\

    \end{array}
  \end{aligned}
\right),
    \label{EqS4}
\end{equation}
meaning only three independent elastic constants: $C_{11}$, $C_{12}$ and $C_{44}$.

\section{SM III. RUS \lowercase{data analysis}}

\begin{figure*}[!ht]
\vspace*{20pt}
\includegraphics[width=15cm]{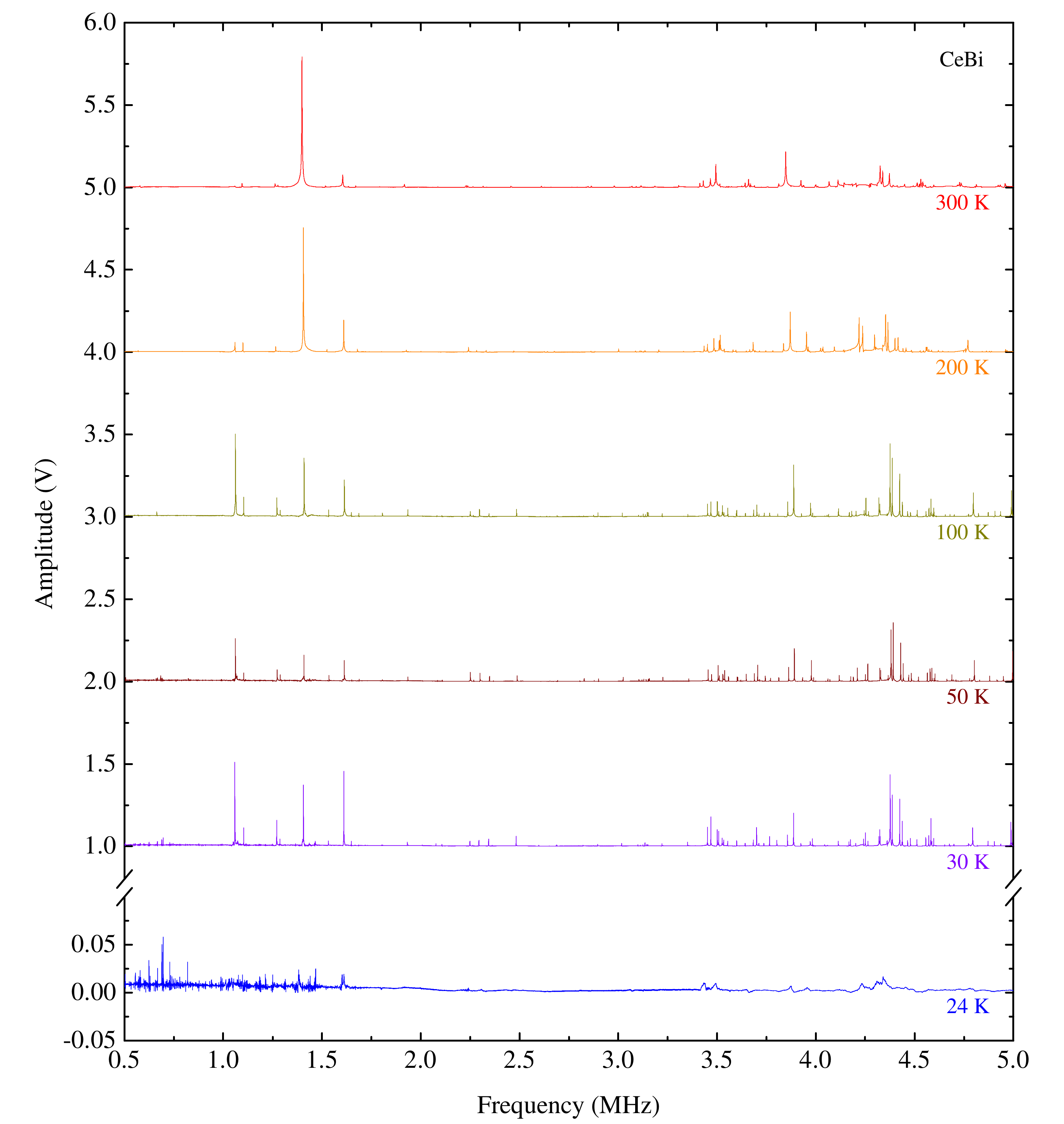}
\vspace*{-0pt}
\caption{\label{FigS2} Resonant ultrasound spectra of CeBi at selected temperatures. The curves have been shifted vertically for clarity. The resonant modes are dissipated severely when approaching $T_{N1}$.}
\end{figure*}

Figure \ref{FigS2} displays the resonant ultrasound spectra of CeBi measured at selected temperatures. The curves have been shifted vertically for clarity. The resonant modes are dissipated rapidly when approaching $T_{N1}$.

If the density $\rho$, symmetry, dimensions and elastic constants $C_{ij}$s of a sample are known, we can calculate the intrinsic vibrational modes by mathematically solving the three dimensions wave function. Reversely, the elastic moduli can also been deduced by the least-square method with a set of known resonant frequencies \cite{Migliori-RUS}. In this fitting, the process can make the initial elastic constant values gradually approach the real values until the root mean square (RMS) $R\equiv\sqrt{\{\sum_{n} [(f^n_{cal}-f_{exp}^n)/f^n_{exp}]^2\}/N}$ minimizes, where $N$ is the total number of recognized peaks, $f_{cal}^n$ and $f_{exp}^n$ are the calculated and experimental frequencies, respectively. Table \ref{table:longtable_example} summarizes the resultant of the analysis for CeBi at 300 K. In our analysis, RMS are less than 0.60\% for all the temperatures between 300 and 30 K.

The uncertainties can be calculated by Levenberg-Marquardt Algorithm \cite{Migliori-RUS}. We expand the $F\equiv\sum_{n=1}^N [(f^n_{cal}-f_{exp}^n)/f^n_{exp}]^2$ to a Taylor series in the independent elastic parameter space $\{C_\alpha\}$, where $\alpha$ = 11, 12 and 44. Combining with the extreme condition $\frac{\partial F}{\partial C_\alpha} = 0$, we can obtain the uncertainties of parameters:
\begin{equation}
    \Delta C_\alpha =  \mathbf{A}^{-1}\mathbf{B},
    \label{EqS5}
\end{equation}
where the elements of tensors $\mathbf{A}$ and $\mathbf{B}$ are:
\begin{equation}
    A_{\alpha\beta} = \sum_{n=1}^N\frac{1}{{f^n_{exp}}^2}\frac{\partial f^n_{cal}}{\partial C_\alpha}\frac{\partial f^n_{cal}}{\partial C_\beta},
    \label{EqS6}
\end{equation}
\begin{equation}
    B_\beta = \sum_{n=1}^N\frac{1}{{f^n_{exp}}^2}\left(f^n_{cal} - f^n_{exp}\right)\frac{\partial f^n_{cal}}{\partial C_\beta}.
    \label{EqS7}
\end{equation}

\setlength{\tabcolsep}{20pt}
\begin{longtable}{cccccccc}

\caption{The fitting results of Resonant ultrasound spectrum of CeBi at 300 K and 0 T. RMS = 0.534 $\%$. Note that Nos. 40, 58 and 63 are ``missing" modes in this measurement.}
\label{table:longtable_example} \\
\hline
\hline
$n$  & $f_{exp}^n$ (MHz)  & $f_{cal}^n$ (MHz)  &  $Error$ ($\%$)  & \multicolumn{3}{c}{$d\ln{f^n_{cal}}/d\ln{C_{ij}}$}
\\\cline{5-7}
    &    &    &     &  $C_{11}$  & $C_{12}$   &  $C_{44}$
\\  \hline
\endfirsthead

\multicolumn{7}{c}%
{{\bfseries \tablename\ \thetable{} -- Continued from previous page}} \\
\hline
\hline
$n$  & $f_{exp}^n$ (MHz)  & $f_{cal}^n$ (MHz)  &  $Error$ ($\%$)  & \multicolumn{3}{c}{$d\ln{f^n_{cal}}/d\ln{C_{ij}}$}
\\\cline{5-7}
    &    &    &     &  $C_{11}$  & $C_{12}$   &  $C_{44}$
\\  \hline
\endhead

\hline \multicolumn{7}{r}{{Continued on next page}} \\ \hline
\endfoot

\hline \hline
\endlastfoot

 1  & 0.569   &  0.556  &  -2.28   &    0.004        &   0.000        &    0.996
\\
  2   & 1.049 & 1.042   & -0.67   & 0.588   &   -0.031  & 0.443
\\
  3   & 1.095 & 1.097   & 0.18   & 0.025    &   -0.000  & 0.975
\\
   4  & 1.262   & 1.256   & -0.48   &  0.257  &  -0.012  & 0.755
\\
   5  & 1.276   & 1.275   &  -0.08  &  0.357   &  -0.015  & 0.658
\\
  6   & 1.399   &  1.389  &  -0.71  & 0.495   &  -0.017  & 0.521
\\
  7   & 1.517   & 1.517   &  -0.00  &  0.355   & -0.021   & 0.665
\\
  8   &  1.605  & 1.601   & -0.25   & 0.164   &  -0.000  & 0.836
\\
  9   &  1.629  &  1.616  & -0.80   & 0.081   & -0.008   & 0.927
\\
  10   & 1.670   & 1.674   & 0.24   &  0.262   &  -0.011  & 0.748
\\
  11   & 1.792   & 1.787   & -0.28   & 0.081  & -0.003     & 0.922
\\
  12   & 1.917   & 1.919   & 0.10   &  0.233  & -0.003   & 0.770
\\
 13    & 1.991   & 1.993   & 0.10   & 0.120   & 0.003   & 0.876
\\
  14   &  2.044  &  2.031  &  -0.64  & 1.088  &  -0.088  & 0.000
\\
  15   &  2.087  & 2.094   & 0.34   & 0.226   &  -0.007  & 0.781
\\
 16   &  2.132  & 2.128   & -0.19   & 0.132   & -0.005   & 0.873
\\
 17   &  2.230  & 2.229   & -0.04   &  0.095  & -0.004   & 0.909
\\
 18    & 2.237   & 2.253  & 0.71  &  0.755  & -0.048   & 0.293
\\
 19    & 2.264   & 2.275   & 0.46   & 1.022   & -0.065   & 0.044
\\
 20    & 2.316   & 2.290   & -1.12  & 0.052   & -0.004   & 0.951
\\
 21    & 2.324   & 2.303   & -0.90   & 0.153   & -0.001   & 0.847
\\
 22    & 2.456   & 2.470   & 0.57   & 0.394   & -0.021   & 0.628
\\
 23    & 2.562   & 2.549   & -0.51   & 1.023   & -0.024   & 0.001
\\
 24    & 2.611   & 2.621   &  0.38  &   0.938  & -0.070   & 0.132
\\
 25    & 2.627   & 2.623   &  -0.15  & 0.225   & -0.006   &  0.781
\\
 26   &  2.641   &  2.636  &  -0.19  & 0.150  &   -0.009  &  0.859
\\
 27   &  2.659   &  2.669  &   0.38 & 0.240   &  -0.005   &  0.765
 \\
 28   &  2.665   &  2.678  &  0.49  & 0.095   &   -0.005  &  0.910
\\
29   &  2.720   &  2.739  &  0.70  & 0.079   &   -0.003  &  0.924
\\
30    &  2.766   & 2.757   & -0.33   &  0.605  &  -0.036   &  0.431
\\
31    &  2.791   & 2.800   &  0.32  & 0.112   & 0.001    &  0.887
\\
 32   &  2.847   &  2.865  &  0.63  & 0.139   &  -0.004   &  0.865
\\
 33   & 2.863    &  2.877  &  0.49  &  0.240  &  -0.016   & 0.776
\\
 34  &  2.979   & 2.981   &  0.07  &  0.463  &  -0.027   &  0.564
\\
35   &  3.066   & 3.072   & 0.20   &  0.678  &  -0.400   &  0.362
\\
36    & 3.069    &  3.082  &  0.42  &  0.746  &  -0.056   &  0.310
\\
37    &  3.091   & 3.100   &  0.29  & 0.294   &  0.000   &  0.706
\\
 38   &  3.105   & 3.101   &  -0.13  & 0.187   & -0.011    &  0.824
\\
 39   &  3.115   &  3.103  &  -0.39  & 0.450   & -0.300    &  0.580
\\
40*    &  -   & 3.124   &  -      &  0.109  &   0.001  &  0.890
\\
 41   &  3.185   & 3.201   & 0.52   &  0.210  & -0.006    &  0.796
\\
 42   &  3.305   &  3.316  & 0.33   & 0.410   & -0.033    & 0.623
\\
 43   &  3.413   &  3.430  &  0.50  &  0.205  & -0.002    &  0.797
\\
44    &  3.430   &  3.462  &  0.93  &  0.196  & -0.008    &  0.812
\\
 45   &  3.466   & 3.468   &  0.06  &  0.198  &  -0.005   &  0.807
\\
46    &  3.488   &  3.488  & 0.00   &  0.189  &   -0.013  &  0.824
\\
 47   &  3.494   &  3.496  & 0.06   &  0.103  &  0.001   &  0.896
\\
 48   &  3.514   &  3.511  &  -0.09  & 0.326   &  -0.017   &  0.591
\\
 49   & 3.550    &  3.526  &  -0.68  &  0.171  & -0.004    &  0.833
\\
 50   &  3.558   &  3.553  &  -0.14  &  0.480  &  -0.030   &  0.551
\\
 51   &  3.574   &  3.585  &  0.31  & 0.212   &   -0.006  &  0.794
\\
 52   &  3.597   & 3.598   &  0.03  &  0.130  &  -0.006   &  0.876
\\
53    &  3.625   &  3.602  &  -0.63  & 0.255   &   -0.010  &  0.755
\\
 54   &   3.642  & 3.637   &  -0.14  &  0.599  &  -0.053   &  0.454
\\
55    &  3.660   & 3.668   & 0.22   &  0.182  &  -0.012   &  0.830
\\
 56   &  3.669   &  3.668  &  -0.03  &  0.215  &  0.009   &  0.794
\\
57    & 3.689    &  3.683  & -0.16  &  0.136  &  -0.005   &  0.869
\\
58*    &  -   &  3.697  & -   &  0.141  &  -0.009   &  0.868
\\
59    & 3.725    &  3.718  & -0.19   &  0.163  &  -0.005   &  0.842
\\
 60   & 3.758    & 3.734   &  -0.64  &  0.573  &  -0.032   &  0.459
\\
  61  &  3.812   &  3.835  & 0.60   &  0.562  &  -0.034   &  0.472
\\
  62  &  3.847   &  3.838  &  -0.23  &  0.229  & -0.010    &  0.782
\\
  63*  & -    &  3.896  &  -  &  0.690  &  -0.022   &  0.332
\\
  64  &  3.874   & 3.897   &  0.59  & 0.251   &  -0.014   &  0.763
\\
  65  &  3.925   &  3.905  & -0.51   & 0.132   &   0.000  &  0.868
\\
  66  &  3.936   &  3.923  &  -0.33  & 0.616   &  -0.042   &  0.426
\\
 67   &  3.939   &  3.973  & 0.86   & 0.130   &   -0.003   &  0.873
\\
 68   & 4.012    &  4.051  & 0.97   & 0.139   &  -0.004   &  0.865
\\
 69   &   4.034  &  4.061  & 0.67   &  0.186  &  -0.007   &  0.821
\\
 70   &  4.050   & 4.070   &  0.49  &  0.210  & -0.012    &  0.802
\\
 71   &   4.067  &  4.075  & 0.20   &  1.080  &  -0.080   &  0.000
\end{longtable}

\end{document}